\documentclass[conference]{IEEEtran}
\IEEEoverridecommandlockouts

\usepackage{cite}
\usepackage{amsmath,amssymb,amsfonts}
\usepackage{graphicx}
\usepackage{booktabs}
\usepackage[ruled,vlined,linesnumbered]{algorithm2e}
\usepackage{array}
\usepackage{siunitx}
\usepackage{subcaption} 
\usepackage{tabularx}
\usepackage{xcolor}

\let\origsubsection\subsection
\renewcommand{\subsection}[1]{\vspace{-0.35em}\origsubsection{#1}\vspace{-0.15em}}
\SetAlFnt{\footnotesize}
\SetAlCapFnt{\footnotesize}
\SetAlCapNameFnt{\footnotesize}

\def\BibTeX{{\rm B\kern-.05em{\sc i\kern-.025em b}\kern-.08em
    T\kern-.1667em\lower.7ex\hbox{E}\kern-.125emX}}

\begin{document}
\title{AoI-Guaranteed Dynamic Route Planning for Connected Vehicles}

\author{%
    Sajedeh Norouzi\textsuperscript{1}, Maryam Ansarifard\textsuperscript{1}, Farshad Zeinali\textsuperscript{1}, Ali Nouruzi\textsuperscript{2}, Nader Mokari\textsuperscript{1},  Hamid Saeedi\textsuperscript{2}, Nizar Zorba\textsuperscript{3} \\
    \textsuperscript{1}Department of Electrical and Computer Engineering, Tarbiat Modares University, Tehran, Iran.\\
    \textsuperscript{2}College of Engineering and Technology, University of Doha for Science and Technology, Doha, Qatar. \\
    \textsuperscript{3} College of Engineering, Qatar University, Doha, Qatar.
    
}

\maketitle
\begin{abstract}
The advancement of Intelligent Transportation Systems (ITS) has been significantly driven by progress in radio communication technology. Dynamic route planning, a key component of ITS, traditionally focuses on metrics such as route capacity and travel time. This paper presents a novel dual-factor approach that integrates travel time estimation and radio resource availability into an innovative route-planning scheme for connected vehicles (CVs).
To address this dual-objective route planning challenge, we employ Deep Reinforcement Learning (DRL). Our approach, called AoI-Guaranteed Dynamic Route Planning (AGDRP), effectively balances travel time and Age of Information (AoI),
\textcolor{black}{enhancing route planning performance through adaptive learning over time.
}
Simulation results demonstrate that AGDRP outperforms the baseline scheme, which solely focuses on travel time optimization. 
\textcolor{black}{In fact, we show that incorporating AoI minimization significantly enhances route planning performance beyond conventional travel-time-based approaches.}
\end{abstract}

\begin{IEEEkeywords}
Intelligent Transportation Systems (ITS), Dynamic Route Planning, V2X Communications, Age of Information (AoI), Deep Reinforcement Learning (DRL).
\end{IEEEkeywords}

\section{Introduction}
\IEEEPARstart{I}{ntelligent} Transportation System (ITS) is an application that brings together state-of-the-art technologies and advanced communication systems with the overarching goal of optimizing transportation networks \cite{app10124314}. A central objective of ITS is to elevate the efficiency, safety, and sustainability of these networks \cite{karimi_mueck_5g_spectrum_2020}. To achieve this, ITS employs cutting-edge technologies, including sensors, cameras, robust communication networks, and sophisticated data analytics. These tools empower ITS to continually gather and analyze real-time data about traffic flow, road conditions, and the overall state of transportation infrastructure  \cite{10026913}. In particular, many researchers use Vehicle-to-Everything (V2X) communication to enhance ITS applications. To improve transportation safety, the authors in \cite{10026913} propose an approach that uses Dedicated Short Range Communication technology to support Intersection Movement Assist (IMA) and Lane Change Assist (LCA). In another work \cite{Wang_Deng_Zhang_Zhang_2020}, the authors use V2X-collected data to calculate resistance values for road sections. In route planning and V2X communication, researchers have championed V2X technologies as valuable resources for acquiring real-time information \cite{oubbati2020search, wevers2017v2x, 8293808}. 

Despite these achievements, inherent limitations in communication systems, such as delay, packet loss, and high Age of Information (AoI), restrict the expected performance of ITS applications, including Green Light Optimal Speed Advisory (GLOSA), safety applications, and dynamic route planning.  A few studies have examined the importance of communication impairment in ITS applications. To deal with this, we can assume two approaches: we can aim to prevent these limitations, such as the work of \cite{10074986}, in which the authors reserve radio and computing resources based on the vehicles' routes to reduce the ratio of tasks not completed before the deadline, which is critical for Connected Vehicles (CVs). 

Alternatively, we can investigate the effects of these limitations and seek to mitigate their drawbacks. The authors in \cite{sharara2019impact} study the impact of packet loss and delay of wireless communication on the performance of GLOSA, where the numerical results show that by increasing the percentage of packet loss, the percentage of arrivals at the green phase decreases, and for lower packet loss values, there is a greater reduction in travel time using GLOSA. Accordingly, the authors propose increasing the activation distance to mitigate the negative effects of packet loss. In another work \cite{8243552}, authors evaluate the systemic performance of CVs' safety due to uncertainties in vehicle position, communication delay, and penetration rate. The effect of AoI on route planning validation was investigated in \cite{norouzi}. By comparing road travel times and volume-over-capacity (V/C) for different levels of information freshness, it was shown that an increase in the average AoI of vehicles degrades route-planning performance.

In this paper, we introduce an innovative dynamic route-planning algorithm to enhance performance. Our approach combines two previously mentioned strategies: mitigating network limitations through efficient use of radio resources and incorporating these considerations into path planning to prevent potential drawbacks.
Firstly, our route selection agent evaluates two key resources when selecting routes: the traditional road capacity \cite{urban-transportation-networks} and the availability of radio resources along those routes \cite{10074986}. 
\textcolor{black}{
Secondly, by optimizing radio resource use, we further mitigate the impact of network limitations. This dual-focus strategy enables more accurate and efficient route planning by leveraging timely network state information.
}
\textcolor{black}{
The efficient use of radio resources aims to minimize the AoI, which captures information freshness and affects the accuracy of network state estimation, rather than directly representing communication reliability, especially for applications that require periodic updates in mesh networks \cite{baldesi2019keep}.
}

For example, if the AoI associated with a road increases, the BS may no longer have an accurate estimate of the road's current state. AGDRP can then discourage assigning additional vehicles to that road until fresher information becomes available. Thus, rather than optimizing travel time alone, AGDRP jointly minimizes AoI and travel time. Reducing AoI improves road-state estimation accuracy, thereby supporting better routing decisions from the user's perspective.
We deploy Deep Reinforcement Learning (DRL) algorithms to enhance route planning by considering diverse resources, thereby reducing decision calculation time and enabling rapid route selection \cite{zeinali2023ai, aboeleneen2024reinforcement}. In particular, we deploy the Deep Deterministic Policy Gradient (DDPG) and Soft Actor-Critic (SAC) algorithms \cite{gharehgoli2023ai} and compare the results.

Our key contributions are as follows. First, we propose a dynamic route-planning framework for CVs that uses V2X updates to account for real-time road conditions. Second, we incorporate AoI into route planning to capture wireless-network limitations and information freshness. Third, we jointly optimize routing and radio resource allocation so that routing decisions account for travel time, traffic management, and real-time communication constraints. Fourth, we compare DDPG and SAC for the joint spectrum access and route selection problem. Finally, we show that AGDRP reduces the average AoI and travel time compared with travel-time-only routing.
% This paper is organized as follows. In the next section, we present the system model and formulate the problem. Then, RL-based solutions are proposed in Section III. In Section IV, we present the simulation results, and Section V concludes the paper.
\section{system model and problem formulation}\label{III}
\subsection{The setup}
We consider an urban case defined in Annex A of \cite{3gpp2016v2x}, composed of $\mathcal{L}=\{1,2, \dots, l, \dots, L\}$ roads, $\mathcal{M}=\{1,2, \dots, m, \dots, M\}$ intersections, $\mathcal{V}= \{1, 2, \dots, v, \dots, V\}$ CVs and a Base Station (BS) located at the center of the area, where $L$, $M$ and $V$ are the total number of roads, intersections and CVs, respectively. We set $v_l(t)=1$ if the CV $v$ is in the road $l$ at time $t$ and $v_l(t)=0$, otherwise. Let $ \mathcal{V}_l(t)= \{1,2, ..., V_l\}$ indicate the set of CVs in road $l$ in each time slot, where $V_l$ denotes the total number of CVs in road $l$. The location of each CV $v$ is denoted by $x_{v}(t)$ at time $t$. 
\begin{figure}
    \centering
\includegraphics[width=0.9\linewidth]{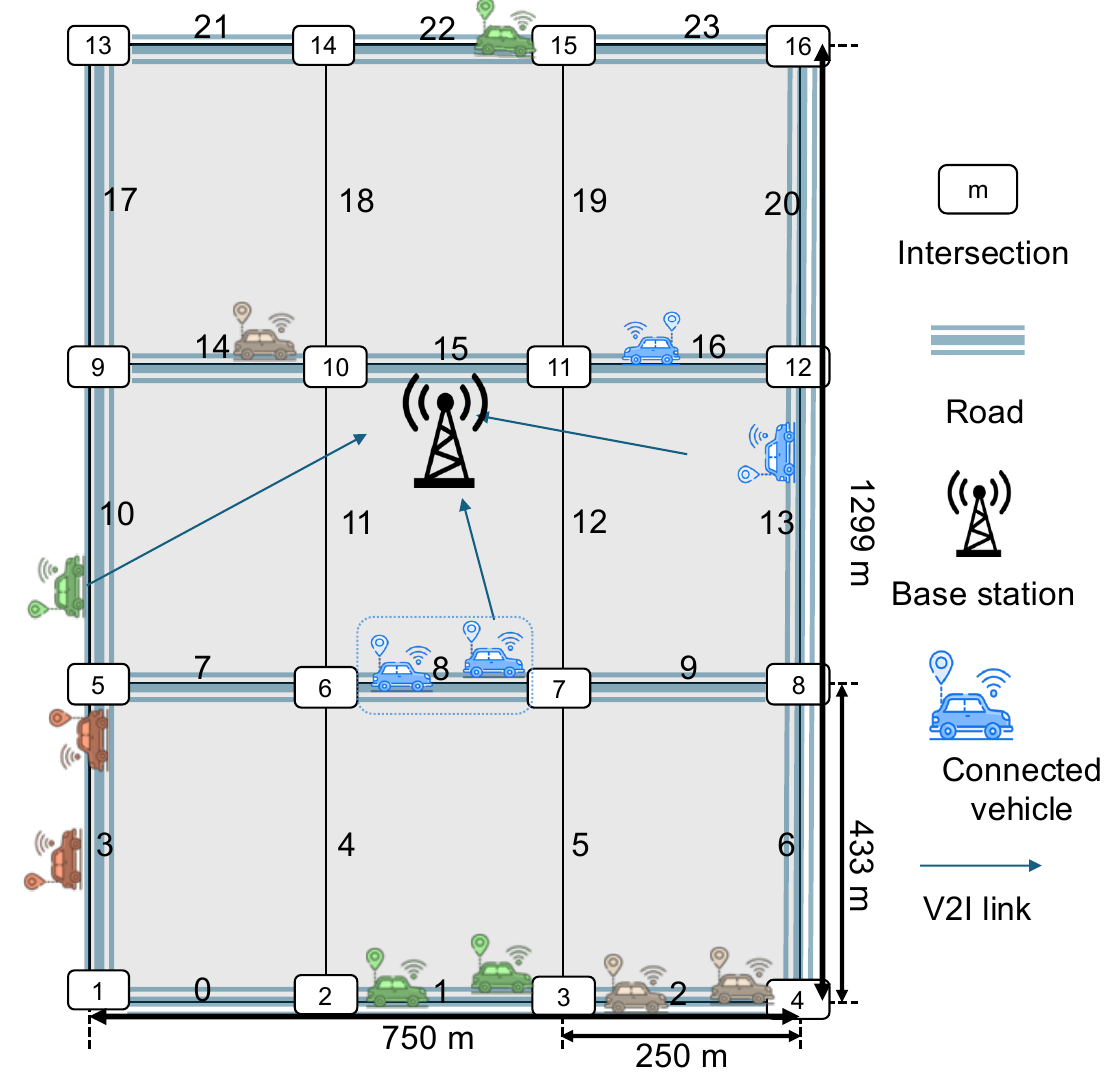}
\caption{Urban road topology from 3GPP Annex A \cite{3gpp2016v2x}.}
    \label{fig:systemmodel}
\end{figure}
Fig.~\ref{fig:systemmodel} illustrates the roads $\mathcal{L}$, intersections $\mathcal{M}$, CVs $\mathcal{V}$, and the central BS that performs routing and communication control. In the route planning framework, when the CV arrives at an intersection, the BS determines an optimal route based on the CV's destination, road capacities, and total travel time. The CV then follows that route. We assume $\delta_{v}(t)=1$ if CV $v$ arrives at its destination, and $\delta_{v}(t)=0$ otherwise. Also, $v_{m}(t)=1$ if CV $v$ reaches intersection $m$ and $v_{m}(t)=0$ otherwise. 
If $v_{m}(t)=1$ and $\delta_{v}=0$, the BS selects the next intersection $b_{v}(t) = m^{\prime}$, where $m^{\prime}\in \mathcal{M}$, $m^{\prime}\neq m$, and only adjacent intersections are eligible.

\begin{figure}[t]
	\includegraphics[width=0.9\linewidth]{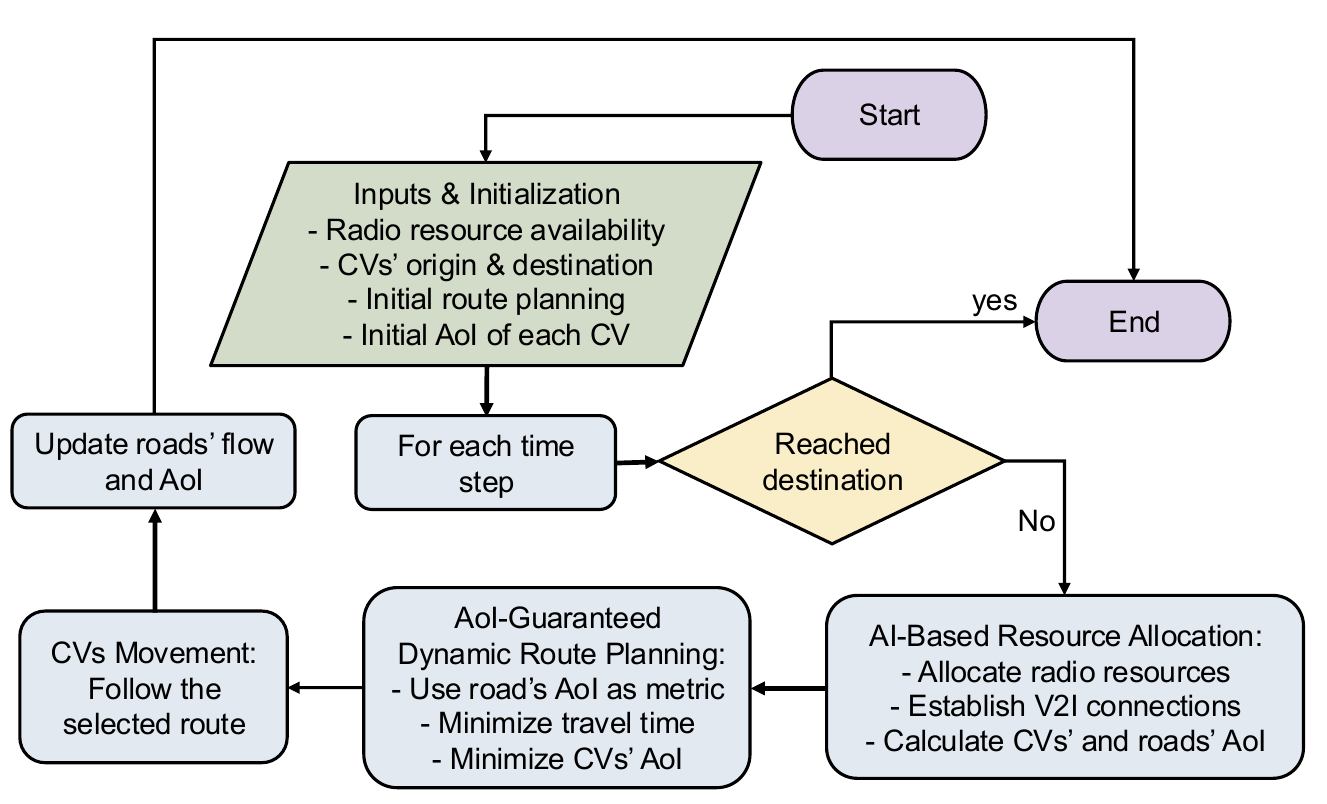}
	\centering
	\caption{\textcolor{black}{AoI-Guaranteed Dynamic Route Planning Flowchart}}
	\label{flowchart} 
\end{figure}

For a CV to send its updated information to the BS, it must be assigned a Resource Block (RB) which is indicated by setting $k_{v}^{n}(t)$ to  $1$ if the RB $n \in \mathcal{N}, \mathcal{N}=\{1, 2, \dots, N\}$ is allocated to CV $v$ at time $t$ and $0$, otherwise, where $N$ is the number of resource blocks. Each resource block can be assigned to at most one CV, 
$\sum_{v=1}^{V}k_{v}^{n}(t)\leq 1 \quad \forall n \in \mathcal{N} $, and each CV can be assigned at most one RB per time slot, $\sum_{n=1}^{N}k_{v}^{n}(t)\leq 1 \quad \forall v \in \{1, \cdots, V\}$. We define the AoI of CV $v$ at time $t$ as the time elapsed since the last time CV had a resource block for exchanging information. The AoI of CV $v$ evolves according to the following recursion:
\begin{align}
	\label{AoI}
	A_{v}(t) = 
	\begin{cases}
		0, &\smallskip\text{if} \; \sum_{n=1}^N k_{v}^{n}(t-1) =1, \\    
		A_{v}(t-1)+1 &\smallskip \text{otherwise}.  \\
	\end{cases}
\end{align}
That is, when CV $v$ is assigned an RB at time $t$, its AoI resets; otherwise, it increases linearly with time. Accordingly, we can calculate the AoI of all CVs located on road $l$ at time $t$, $A_{l}(t)$, which indicates road condition awareness as
\begin{align}
	\label{aoiroad}
	A_{l}(t) = \sum_{v=1}^{V}  v_l(t)A_{v}(t), \forall t.
\end{align}
Travel time serves as a fundamental metric in route planning. The pursuit of optimal route planning to minimize travel time and circumvent congested routes has garnered significant attention from academic researchers and traffic management organizations \cite{8293808}. We calculate the travel time of road $l$ from the Bureau of Public Roads (BPR) as follows:
\begin{align}\label{traveltime}
	T_l = ft  \left[1 + \alpha \left(\frac{F_l(t)}{C_l(t)}\right)^{\beta} \right],
\end{align}
where $ft$ is the free flow travel time and, $\alpha$ and $\beta$ are adjusting parameters. Also, $F_l(t)$, $C_l(t)$ are the flow and capacity of road $l$, respectively. Accordingly, the flow of each road is defined as the number of CVs on it at a given instant, $t$. Due to the dynamic conditions of the roads and incidents, we can not assume a fixed value for capacity. Instead, we model it as follows:
\begin{align}
	\label{capacity}
	C_l(t) = \hat{C}_l(t) + \Delta C_l(t),
\end{align}
where $\hat{C}_l(t)$ and $\Delta C_l(t)$ denote the estimated capacity of each road and its deviation from the actual value, respectively. The deviation term $\Delta C_l(t)$ is modeled as a noise component whose magnitude depends on the Age of Information (AoI), reflecting the reduced accuracy of outdated information. In particular, larger $A_l(t)$ leads to higher estimation error, whereas smaller $A_l(t)$ yields more accurate capacity estimation.
\footnote{In our simulations, we set $\hat{C}_l(t) = C_{\text{max}, l} - \hat{F}_l(t)$, where $\hat{F}_l(t)$ represents the flow on road $l$ derived from the data collected from the CVs. Due to network limitations and the lack of real-time updates, this estimated flow may deviate from the actual value. This inaccuracy is incorporated into the simulation by using the normalized average flow of all CVs on the road, weighted by a factor.}
The proposed AoI-guaranteed route-planning algorithm is presented in Algorithm~\ref{algorithm}, and its overall workflow is illustrated in Fig.~\ref{flowchart}.
\subsection{Problem Formulation}
%In this section, we formulate an optimization problem for AoI-guaranteed dynamic route planning. 
At the beginning of each time step, the intelligent controller in the BS selects the direction for each CV to balance route-planning and V2X communication performance.
Our objective is to minimize the average travel time of CVs over the considered period of $T$ seconds while also minimizing their AoI. The optimization variables are resource block allocation $k_{v}^{n}(t)$ and intersection selection. The multi-objective optimization problem is formulated as
\begin{align}		\label{objective function} 
	\begin{aligned}
		&\min _{\boldsymbol{k},\boldsymbol{b}}\left\{\frac{1}{L}\sum_{l=1}^{L} T_{l}, \sum_{l=1}^L\sum_{t=1}^T A_{l}(t)\right\}   \\
		%\text{ s.t.} \: & C 1: \sum_{v_l=1}^{V_l} v_L^{t} = F_l(t), \quad \forall l \in \mathcal{L}, \forall t \in {M}, \\
		\: &C 1: \sum_{n=1}^{N}k_{v}^{n}(t)\leq 1, \quad \forall v \in \mathcal{V}, \forall t, \\
		\: &C 2: \sum_{v=1}^{V}k_{v}^{n}(t)\leq 1, \quad \forall t, \\
		\: &C 3: \delta_{v} = \{0, 1\}, \quad \forall v\in \mathcal{V},\\		
		\: & C 4: v_{m}(t) = \{0, 1\}, \quad \forall m \in \mathcal{M}, \forall t,\\
	\end{aligned}
\end{align}
where, $\boldsymbol{k}=\{k_{v}^{n}(t)|v\in \mathcal{V}, n\in  \mathcal{N}, \forall t\}$ and $\boldsymbol{b}=\{b_{v}(t)|v\in \mathcal{V}, \forall t\}$. The problem is a mixed binary, nonlinear, and non-convex optimization problem that is not tractable. In this regard, we utilize two DRL algorithms, single-agent DDPG and SAC, to address the complexity of the proposed optimization problem.  
% \begin{algorithm}
% 	\scriptsize
% 	%\tiny
% 	\renewcommand{\arraystretch}{0.4}
% 	\caption{AoI Guaranteed Dynamic Route Planning}
% 	\label{algorithm}
% 	Establish the urban environment, including the length of the road network and the locations of intersections. \\
% 	Define the origins and destinations of CVs. \\
% 	Determine road capacities and compute the free travel time for each road segment.
% 	Set up a radio environment, including V2I links, radio resources, and the location.
% 	\While{All CVs reach their destinations}{
% 		\For{each time step}{
% 			Resource blocks allocation through DRL algorithm \\
% 			Calculate AoI of all CVs (\ref{AoI}) and roads (\ref{aoiroad})\\
% 			Determine travel time and capacity for each road by analyzing data collected from CVs (\ref{traveltime}), (\ref{capacity})\\
% 			Consider the impact of network limitations on the accuracy of the estimation.\\ %(\ref{capacityerror}). \\
% 			\For{all CVs}{
% 				\If{CV reaches intersection}{
% 					Route planning (intersection selection to move forward to) through the DRL algorithm by considering:
% 					\\ \indent Road capacity, the AoI of optional roads 
% 					\\ \indent Travel time for each CV 
% 				}
% 				\Else{CV follows the previous route} 
% 			}
% 		}
% 	}
% \end{algorithm}
\begin{algorithm}[t]
\footnotesize
\caption{AoI-Guaranteed Dynamic Route Planning}
\label{algorithm}
\KwIn{Urban road network (segments, intersections), origins and destinations of CVs, road capacities, radio resources, V2I coverage map}
\KwOut{Dynamic routes for all CVs with AoI-aware updates}
\BlankLine
\textbf{Initialization:}\;
    Establish the urban environment (road lengths, intersection locations)\;
    Define the origin--destination (OD) pairs of all connected vehicles (CVs)\;
    Compute the free-flow travel time for each road segment\;
    Configure the radio environment: V2I links, available resource blocks, and BS locations\;
\BlankLine
% \For{there exists at least one CV that has not reached its destination}{
  \For{each time step $t$}{
    % \tcp{Radio resource management}
    Allocate resource blocks via the DRL-based RRM scheme\;
    % \tcp{State estimation from CV measurements}
    Compute the AoI of all CVs using \eqref{AoI}\;
    Compute the AoI of all roads using \eqref{aoiroad}\;
    Update estimated travel time and capacity of each road segment using \eqref{traveltime} and \eqref{capacity}\;
    Account for the impact of network limitations on estimation accuracy\; % (e.g., delayed or missing updates)
    % \tcp{AoI-aware route planning at intersections}
    \For{each CV $v$}{
      \If{CV $v$ reaches an intersection}{
        Select the next outgoing road (route decision) via the DRL-based planner by considering:\;
        \Indp
        road capacity estimates,\;
        AoI of candidate outgoing roads,\;
        and predicted travel time for CV $v$\;
        \Indm
      }
      \Else{
        CV $v$ continues along its previously selected route\;
      }
    }
  }
\end{algorithm}

\section{Solution: DDPG and SAC}\label{V}
 \subsection{General Setup}
We first describe the Markov decision process (MDP) model, which is defined as a tuple $\left\{\mathcal{S}_t,\mathcal{A}_t,\mathcal{R}_t,\mathcal{S}_{t+1}\right\}$ containing states, actions, rewards, and new states. The agent collects state information during the DRL process to determine the scheduling actions. The information includes CVs' locations, selected intersections, AoI of all roads, CVs' arrival, capacity of each road, and travel time of roads, which is expressed as
\begin{align}\label{state}
	s_t = \left\{x_{v}(t), b_{v}(t), A_{l}(t), \delta_{v}(t), C_l(t), T_l\right\}.
\end{align}
The agent decides what action to take based on the network's current state as it transitions to a new one. The action space in our problem consists of selecting points for CVs and channel assignments. Accordingly, $a_t \in \mathcal{A}_{t}$ executed in time step $t$, is given as
%\begin{align}\label{actions}
$	a_t = \left\{k_{v}^{n}(t), b_{v}(t)\right\}.$
%\end{align} 
\textcolor{black}{
Although the actor network outputs continuous actions, these values are mapped to discrete routing and binary channel–allocation decisions within the environment. Specifically, the next intersection is selected by taking an $\arg\max$ over candidate actions. For channel allocation, each CV is assigned the resource block with the highest output value, while ensuring that each resource block is allocated to at most one CV. Thus, the learning agent operates in a continuous action space, while the underlying control decisions are discrete. After an action is executed, the agent receives an immediate reward, denoted as
\begin{align}
    r_t = -w\left(\frac{1}{L}\sum_{l=1}^{L} T_l\right) - w'\left(\frac{1}{L}\sum_{l=1}^{L} A_l(t)\right)
\end{align}
}
where $w$ and $w'$ are weighting coefficients, $T_l$ represents the travel time on road $l$, $A_l(t)$ denotes the AoI associated with road $l$ at time $t$, and $L$ is the total number of road segments considered. The negative sign encourages the agent to minimize both average travel time and cumulative AoI. 
The long-term accumulated reward of a policy is modeled as 
%\begin{align}
$	R_{t} = \sum_{i=0}^{\infty}\gamma^{i}r_{t+i},$
%\end{align}
The discount factor $\gamma \in \left[0, 1\right]$ indicates that the agent is more concerned about the long-term reward if the discount factor approaches $1$.

\subsection{Single Agent DDPG}
The DDPG algorithm leverages the advantages of the Deep Q-Network (DQN) and the actor-critic algorithm, which employ two Deep Neural Networks (DNNs), an actor and a critic, to approximate the policy function and the value function, respectively. Actor $\mu(s|\theta^{\mu})$ with weights $\theta^{\mu}$, makes action selection, and critic $Q(s, a|\theta^{Q})$ with weights $\theta^{Q}$, evaluates the policy function. Target networks $Q^{\prime}$ and $\mu^{\prime}$ are also employed to avoid divergence of the learning algorithm due to direct updates of the network weights based on Twin Delay (TD) error gradients. The value network is updated based on the Bellman equation by minimizing the mean-squared loss between the updated Q value (see \eqref{y_j} below) and the original value, which can be formulated as 
\begin{align}\label{y_j}
	y_{j} = r_{j}+\gamma Q^{\prime}(s_{j+1}, \mu^{\prime}(s_{j+1}|\theta^{\mu^{\prime}})|\theta^{Q^{\prime}}),
\end{align}
\begin{align}
	L = \frac{1}{N_{\text{Tran}}}\sum_{j}(y_{j}-Q(s_j, a_j|\theta^{Q}))^2,
\end{align}
where $\{s_j, a_j, r_j, s_{j+1}\}$ is a sampled random batch of $N_{\text{Tran}}$ transitions. Based on the deterministic policy gradient theorem, the policy network’s update is calculated as follows:
\begin{align}\label{actor update}
	\Delta_{\theta^{\mu}}\mu|_{s_j} \approx \frac{1}{N_{\text{Tran}}} \sum_{j}\Delta_{a}Q(s_j, a|\theta^{Q})|_{a=\mu(s_j)} \Delta_{\theta^{\mu}}\mu(s_j|\theta^{\mu}).
	\nonumber
\end{align}
Finally, the target networks are updated with the learning rate of $\alpha_{\text{lr}}$ for the actor and $\beta_{\text{lr}}$ for the critic networks as\\ %\begin{align}\label{targetupdate}
$	\theta^{Q^{\prime}} \leftarrow \alpha_{\text{lr}} \theta^{Q} + (1-\alpha_{\text{lr}}) \theta^{Q^{\prime}}$
and 
$	\theta^{\mu^{\prime}} \leftarrow \beta_{\text{lr}} \theta^{\mu} + (1-\beta_{\text{lr}}) \theta^{\mu^{\prime}}.$
%\end{align}

\subsection{Single Agent SAC} 
SAC is a DRL algorithm for continuous action spaces, based on the principle of maximum entropy. The objective of this approach, however, is to optimize a reward that is regularized by entropy, rather than to maximize the cumulative discounted reward. Then, the following formula maximizes the entropy:
\begin{equation}
	J(\mu)=\sum_{t=0}^T \mathbb{E}_{\left(s_t, a_t\right) \sim \rho_\mu}\left[r_t+\varphi \mathcal{H}\left(\mu\left(\cdot \mid s_t\right)\right)\right].
\end{equation}
The regularization coefficient $\varphi$ in this equation ensures that entropies and the sum of expected rewards remain finite. For infinite-horizon problems, the SAC method extends soft policy iteration when using function approximation. Instead of estimating the true $Q$ value of policy $\mu$ to enhance it, SAC adopts a different optimization approach for both the policy and the value function. We use the same actor and critic networks as in DDPG. The $Q$ function is learned by minimizing the soft Bellman residual, as follows:
\begin{align}
	\nonumber
	& J_Q(\theta^Q)= 
	 \mathbb{E}\left[\left(Q\left(s_t, a_t\right)-r_t-\gamma \mathbb{E}_{s_{t+1}}\left[V_{\theta^{Q'}}\left(s_{t+1}\right)\right]\right)^2\right], \\
	& V_{\theta^{Q'}}(s)= \mathbb{E}_{\mu_{\theta ^ \mu}}\left[Q_{\theta^{Q'}}(s, a)-\varphi \log \mu_{\theta^\mu}(a \mid s)\right].
\end{align}
We can \textcolor{black}{also learn} the policy by minimizing the expected KL-divergence $\mu_{\theta^\mu}$:
\begin{equation}
	\label{54}
	J_\mu(\theta^\mu)=\mathbb{E}_{s \sim }\left[\mathbb{E}_{a \sim N_{\text{Tran}} \mu_{\theta ^ \mu} }\left[\varphi \log \mu_{\theta^\mu}(a \mid s)-Q_{\theta^Q}(s, a)\right]\right],
	%\nonumber
\end{equation}
$N_{\text{Tran}}$ represents the set of previously sampled states and actions, typically stored in a replay buffer. SAC employs two Q-networks along with two target Q-networks to mitigate the issue of biased Q values, denoted as $Q_{\theta^Q}(s, a)=\min \left( Q_{\theta^{Q'}}(s, a), Q_{\theta^{Q''}}(s, a) \right)$. There are various ways to optimize $J_\mu(\theta^\mu)$. Still, one effective approach is to use a likelihood-ratio gradient estimator, which eliminates the need to back-propagate gradients through the target density networks and the policy. In the case of SAC, where the target density is represented by the Q-function, we can employ the reparameterization trick for the policy network, yielding a lower-variance estimator. To achieve this, we re-parameterize the policy $\mu_{\theta^\mu}$ using a neural network transformation that takes both the state $s$ and a noise vector $\epsilon$ as inputs, as follows:
\begin{equation}
	\label{55}
	a=f_{\theta^\mu}(s, \epsilon) .
\end{equation}
Accordingly, the following can be derived using  \eqref{54} and \eqref{55}:
\begin{equation}
	\begin{aligned}
		& J_\mu(\theta^\mu)= \\
		&\mathbb{E}_{s \sim N_{\text{Tran}}, \epsilon \sim \mathcal{N}}\left[\varphi \log \mu_{\theta^\mu}\left(f_{\theta^\mu}(s, \epsilon) \mid s\right)-Q_{\theta^Q}\left(s, f_{\theta^\mu}(s, \epsilon)\right)\right],
	\end{aligned}
	\nonumber
\end{equation}
where $\mathcal{N}$ represents a standard Gaussian distribution, and $\mu_{\theta^\mu}$ is implicitly defined in terms of $f_{\theta^\mu}$. Lastly, SAC provides an approach to automatically update the regularization coefficient $\varphi$ by minimizing the loss function as follows:
\begin{equation}
	J(\varphi)=\mathbb{E}_{a \sim {\mu _{\theta ^\mu} }}\left[-\varphi \log \mu_{\theta^\mu} (a \mid s)-\varphi \mathfrak{e}\right],
\end{equation}
where $\mathfrak{e}$ represents the target entropy as a hyperparameter.
 
\section{Simulation results}\label{VII}
\begin{figure}[t]
	\includegraphics[width=1\linewidth]{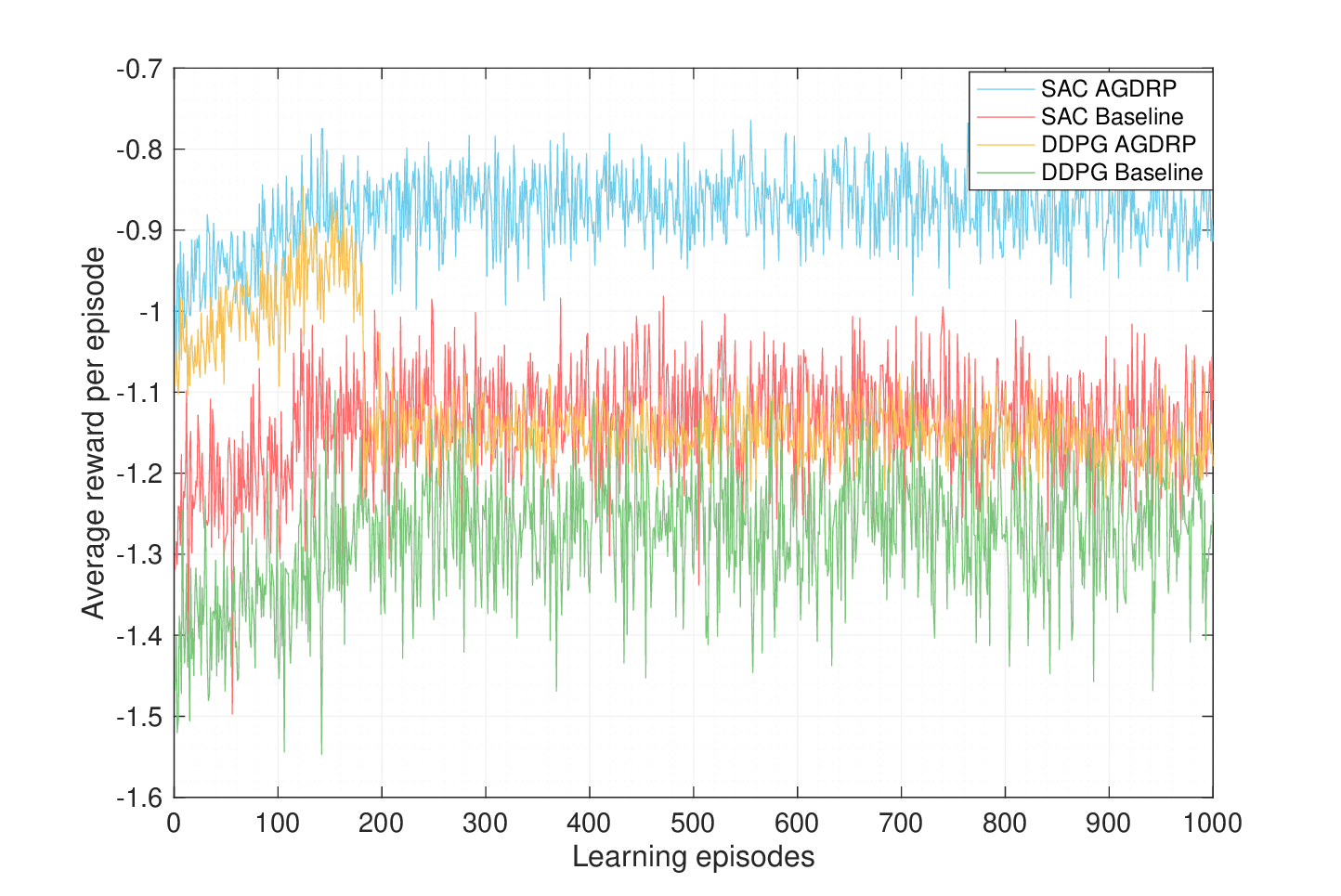}
	\centering
	\caption{Reward of single-agent DDPG and SAC‌ algorithms over learning episodes.}
	\label{reward} 
\end{figure}
We evaluate the proposed framework on the urban network shown in Fig.~\ref{fig:systemmodel}, comprising 16 intersections, 24 bidirectional roads, 40 CV origin-destination pairs, and a central BS. The BS acts as the control agent: it receives CV updates, estimates link conditions, allocates radio resources, and selects the next intersection for each CV. The main simulation parameters are summarized in Table~\ref{table:simulation_parameters}. The evaluation platform is a discrete-time system-level simulator. It jointly captures vehicle mobility, AoI evolution, link flow, capacity estimation, and BPR-based travel time. DDPG and SAC are implemented with replay buffers and target-network soft updates. For reproducibility, all episodes use the same fixed random seed and randomly generated CV origin-destination pairs. At each decision step, the DRL state includes CV locations, destinations, selected intersections, AoI values, arrival flags, link flows, and accumulated travel times; the continuous actor output is then mapped to feasible next-intersection and RB/subchannel decisions. We compare AGDRP with a travel-time-only baseline obtained by removing the AoI term from \eqref{objective function}.
%For simulations, we analyzed an urban environment comprising 16 intersections, 24 two-way roads, 40 origin-destination pairs, and a central BS. Detailed simulation parameters are provided in Table \ref{table:simulation_parameters}. The BS acts as the agent, allocates subchannels to CVs, estimates road capacity and travel time from received updates, and determines routes. The system-level evaluation is implemented as a discrete-time Python simulator. The environment module models road topology, vehicle mobility, AoI evolution, link flow, capacity estimation, and BPR-based travel time; DDPG and SAC are implemented in PyTorch with replay buffers and target-network soft updates. At the beginning of each episode, CV origin-destination pairs are randomly generated over the 16 intersections using a fixed seed. The state concatenates normalized CV locations, destination coordinates, selected next intersections, per-CV AoI, destination-arrival flags, link flows, and accumulated travel times. The DRL action vector is mapped to discrete next-intersection choices and RB/subchannel assignments under the adjacency and allocation constraints. The simulator saves reward, AoI, road-level travel time, end-to-end CV travel time, and destination-arrival traces as \texttt{.mat} files for figure generation. As already explained, we consider a baseline scenario in which, in \eqref{objective function}, we only minimize travel time and remove the second term, i.e., the average AoI. 
%The code used for these simulations is available in \cite{Norouzi2024}.
Fig.~\ref{reward} shows the convergence behavior of the DDPG and SAC frameworks for our proposed AGDRP algorithm and the baseline method. When solved by SAC, our proposed AGDRP framework demonstrates more stable performance and achieves higher rewards than DDPG. The performance of DDPG, although better than the baseline scheme, suggests that it may be inefficient in addressing dual-factor problems, such as the one in our case. 

\begin{figure}
    \centering
    \includegraphics[width=0.85\linewidth]{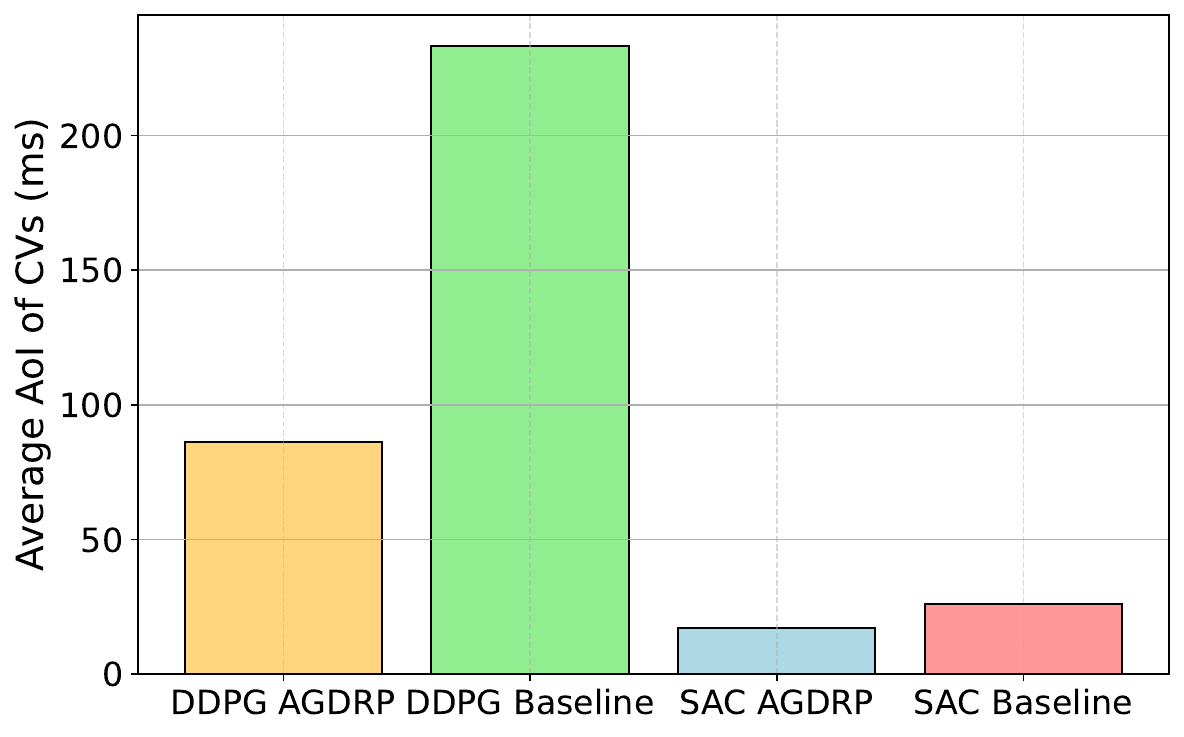}
    \caption{Average AoI of CVs (ms), computed over all 40 CVs.}
    \label{fig:aoi}
\end{figure}

\begin{figure}
    \centering
    \includegraphics[width=0.85\linewidth]{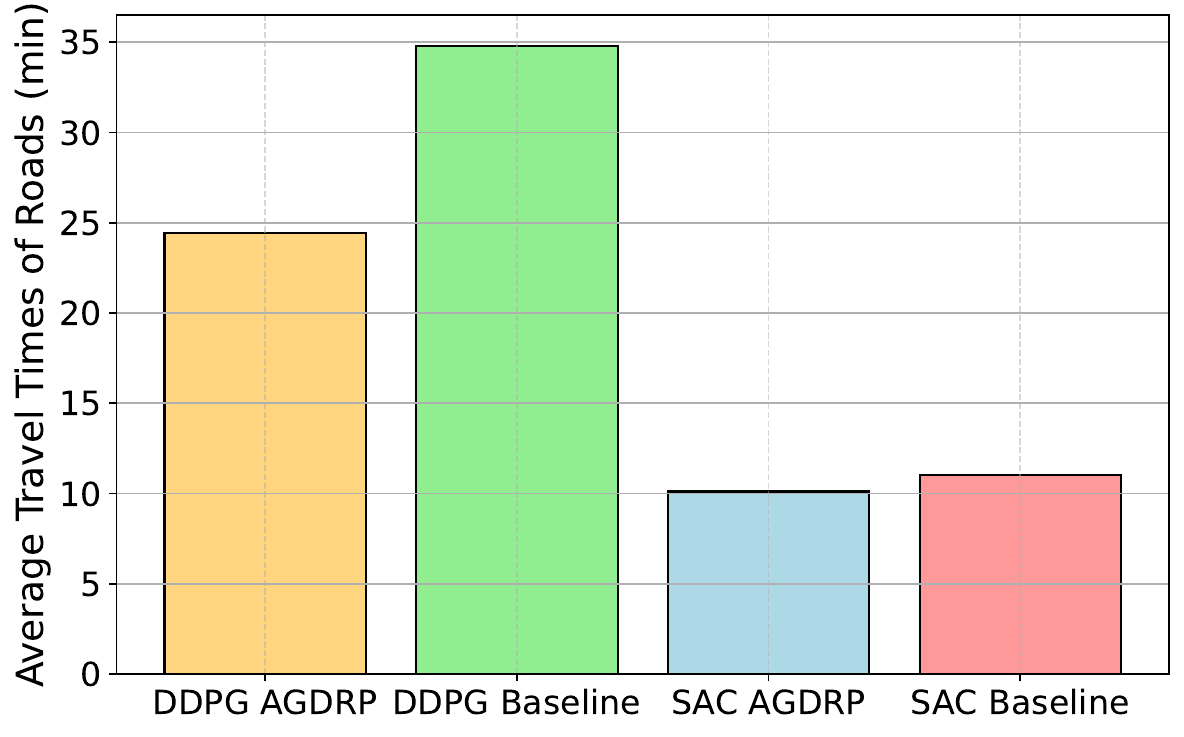}
    \caption{Average travel time on roads (min), computed over all 24 roads.}
    \label{fig:ttroad}
\end{figure}

To compare the algorithms, we evaluate both the communication and mobility metrics across the simulated road network. Fig.~\ref{fig:aoi} presents the average AoI of all CVs at a specific time step. SAC-AGDRP achieves the lowest AoI, consistently maintaining values below 50 ms. AGDRP also performs notably well, achieving lower AoI values than its corresponding baseline in both the DDPG- and SAC-based implementations. Meanwhile, SAC generally outperforms DDPG in terms of communication freshness.
Fig.~\ref{fig:ttroad} shows the average travel time on roads, computed as the time required for vehicles to traverse each road under different routing strategies. This metric captures road-level mobility behavior and provides a finer-grained assessment of how well each algorithm distributes traffic across the 24 road segments, a key indicator of congestion management and load balancing. The results indicate that AGDRP outperforms the baseline algorithm in reducing road‐level travel times across both DRL frameworks.

\begin{figure}
    \centering
    \includegraphics[width=0.85\linewidth]{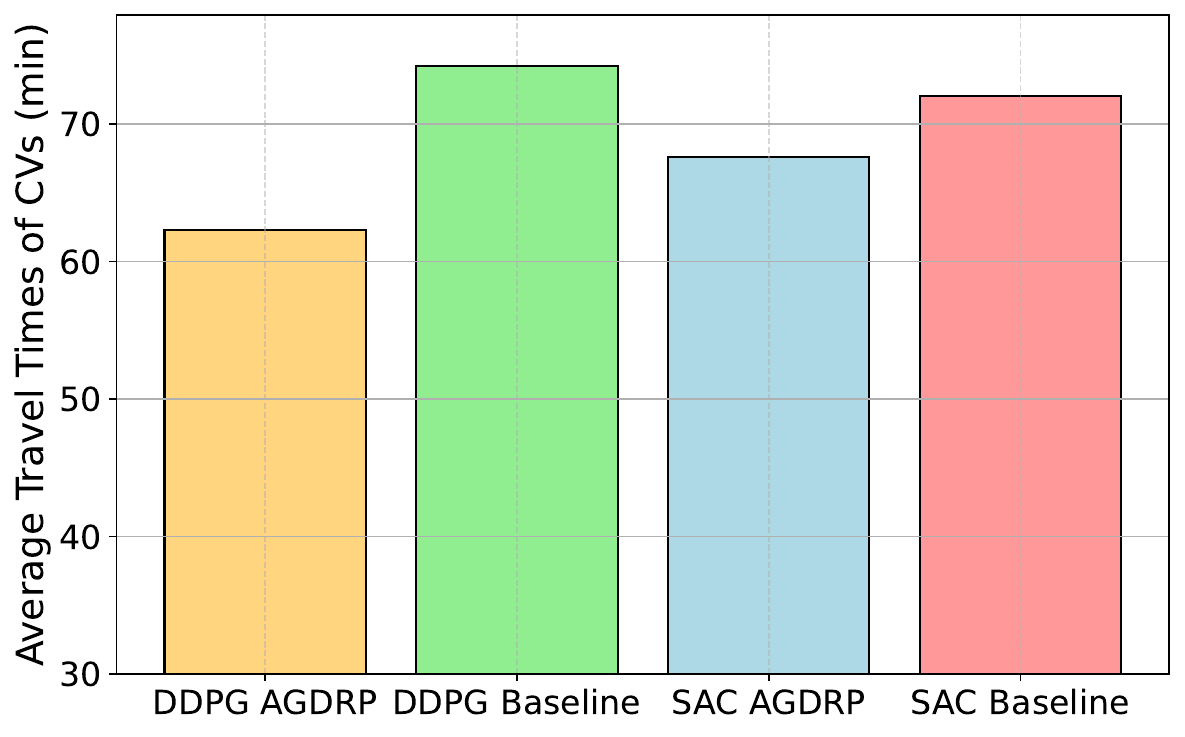}
    \caption{Average travel time (min) across all CVs, including complete journeys.}
    \label{fig:ttvehicle}
\end{figure}

\textcolor{black}{
Finally, Fig.~\ref{fig:ttvehicle} illustrates the average end-to-end travel time of CVs from departure to destination. Consistent with the previous results, AGDRP achieves shorter travel times under both the DDPG and SAC algorithms. When combined with the AoI results in Fig.~\ref{fig:aoi}, these findings highlight that SAC-AGDRP provides a balanced advantage by jointly improving mobility and information freshness. In particular, lower AoI leads to more accurate estimates of road conditions, enabling more informed routing decisions. In contrast, the baseline algorithm operates with less up-to-date information, resulting in less efficient route selection and higher travel times. Overall, AGDRP leverages timely information to enhance both routing accuracy and mobility performance.}
This evaluation can be extended by varying the number of vehicles, the distribution of origins and destinations, and the network scale to assess the robustness of the proposed approach under different traffic densities and routing conditions.
\begin{table}[t]
\caption{Simulation Parameters}
\footnotesize
\label{table:simulation_parameters}
\centering
\begin{tabular}{llc}
\hline
\textbf{Parameter} & \textbf{Notation} & \textbf{Value} \\
\hline
\multicolumn{3}{l}{\textbf{Environment Parameters}} \\
Num of CVs, subchannels & $V$, $N$  & 40, 10 \\
Num of roads, intersections & $L$, $M$& 24, 16\\
Road capacity,  & $C_l(t)$ & 4 \\
Road lengths & -- & 250, 433 m \\
Free-flow time & $ft$ & 18, 31 s \\
CV speed range & -- & 5--10 m/s \\
Time slot duration & -- & 1 ms \\
Initial AoI & $A_v(0)$ & 5 ms \\
BPR parameters & $\alpha,\beta$ & 0.15, 4 \\
Replay buffer size & -- & $10^{6}$ \\
Mini-batch size & $N_{\mathrm{Tran}}$ & 32 \\
Discount factor & $\gamma$ & 0.9 \\
Soft-update coefficient & $\tau$ & 0.01 \\
Training episodes & -- & 1000 \\
Training steps/episode & -- & 1000 \\
\hline
\multicolumn{3}{l}{\textbf{SAC/ DDPG Parameters}} \\
SAC Actor hidden layers & -- & [1024, 512]\\
SAC Critic hidden layers & -- & [1024, 512, 256]\\
SAC Value hidden layers & -- & [256, 256] \\
DDPG Actor/ Critic hidden layers & -- & [400, 300], [400, 300] \\
\hline
\end{tabular}
\end{table}
To evaluate the computational efficiency of the proposed algorithms, we analyze the complexity of both SAC-AGDRP and DDPG-AGDRP. In each time step, both methods require forward passes through their actor and critic networks, and the computational cost scales with the number of network parameters. For DDPG-AGDRP, the per-step complexity is dominated by one actor network and one critic network, yielding a computational cost of 
$\mathcal{O}(P_{\text{actor}} + P_{\text{critic}})$. 
SAC-AGDRP, in contrast, employs a stochastic actor, two critic networks, and an additional value network, resulting in a higher per-step complexity of 
$\mathcal{O}(P_{\text{actor}} + 2P_{\text{critic}} + P_{\text{value}})$. 
During training, both algorithms perform mini-batch updates sampled from a replay buffer of size $10^6$, which introduces a batch-update cost that scales linearly with the mini-batch size. 
Overall, while SAC-AGDRP requires more computation than DDPG-AGDRP, the improved stability and exploration provided by SAC translate into superior performance in both AoI and mobility metrics, demonstrating that the increased computational cost is justified by the performance gains.
The main cost of AGDRP in larger networks is the growth of the state and action spaces. The state dimension scales with the number of CVs and road segments, while the action dimension scales with the number of CVs and RBs/subchannels, since the BS jointly determines routing and resource allocation. Hence, although online inference with a trained policy is fast, training can be computationally expensive, and centralized operation may add signaling overhead. The current evaluation also uses a single-BS grid and an AoI-based capacity-estimation model. Larger deployments with multiple BSs, heterogeneous demand, packet errors, handovers, and realistic wireless channels may require hierarchical control, policy decomposition, or multi-agent DRL. These limitations motivate further city-scale scalability studies.
\vspace{-4 mm}
\section{conclusion}\label{VIII}
In this paper, we propose a dynamic route-planning framework that jointly accounts for road capacity constraints and radio resource availability to optimize CV routes. By integrating these two dimensions, the framework effectively minimizes both travel time and AoI, thereby addressing the critical interplay between mobility and communication constraints in vehicular networks. The DRL-based solution simultaneously performs AoI-aware resource allocation and adaptive routing, enabling vehicles to make informed decisions based \textcolor{black}{on timely and accurate information.}
We benchmark the proposed method against a baseline algorithm that focuses solely on minimizing travel time, as is common in traditional routing strategies. Simulation results show that the SAC-based implementation consistently outperforms its DDPG counterpart, delivering notable improvements in both communication freshness and mobility efficiency. Overall, the proposed AGDRP framework represents a significant advancement in communication-aware route optimization, highlighting the importance of incorporating network information into ITS decision-making processes.
Future work includes scaling AGDRP to larger, more realistic networks and integrating richer communication models to better reflect real-world conditions. Exploring multi-agent learning, predictive mobility/communication models, and deployment in digital twins or testbeds will further advance its practical applicability.
\vspace{-3 mm}
\bibliographystyle{IEEEtran}

\bibliography{Citation}
\end{document}